# Elastic Interferometers as Phononic Filters for Thermal Isolation in Low-Noise Transition Edge Sensors

Djelal Osman[*], Stafford Withington, David J. Goldie and Dorota M. Glowacka

*University of Cambridge, Cavendish Laboratory, JJ Thomson Avenue, Cambridge, CB3 0HE*

*Contact: d.osman@mrao.cam.ac.uk

*Abstract*—**In order to achieve low Noise-Equivalent-Powers (NEP < 1 aW Hz$^{-1/2}$) Transition Edge Sensors (TES) require high levels of thermal isolation between the superconducting bilayer and the heat bath. We propose that short micro-machined acoustic interferometers can be used for low-noise thermal isolation, avoiding many of the difficulties inherent in conventional, long-legged TES designs. In this paper, we present a detailed elastic wave model of interferometric phononic legs, and demonstrate the successful fabrication of TESs that are thermally isolated by interferometric structures.**

## I. INTRODUCTION

Low-noise Transition Edge Sensors (TES) require very low thermal conductances between the superconducting bilayer and heat bath. To date, the lowest conductances have been achieved by using long, narrow, amorphous $Si_xN_y$ support legs, and relying on diffusive transport to bring about a 1/L dependence, where L is the length of the support leg [1,2,3,4]. In fact, the conductance scales as 1/N, where N is the number of acoustic attenuation lengths along the physical length of the structure. To achieve ultra-low Noise Equivalent Powers (NEPs) requires long, narrow legs, which brings numerous difficulties: (i) It is not possible to densely pack pixels into imaging arrays, and so light pipes are needed, but it is extremely difficult to produce long, high optical efficiency light-pipes at wavelengths of tens of microns. (ii) The legs become frail, yields are potentially reduced, and there is concern that the legs may not survive the launch of a space instrument. (iii) Several groups report conductance variations of around ±15% from nominally identical devices on the same wafer, and we believe that this may be due to resonant phonon localization caused by disorder. The disorder that is responsible for the 1/L dependence in the first instance, leads to resonant phonon trapping when the cross section of the microbridge is made exceedingly small. (iv) The heat capacity of the legs can have a marked effect on the electro-thermal characteristics of a TES, and therefore low-heat-capacity legs are beneficial. Various efforts have been made towards controlling conductances by alternative means. Some have included roughening the surface of crystalline Si [5], but most have been based on large-scale patterning [6,7,8], with varying degrees of predictability. In this paper we propose that micro-machined elastic interferometers can be used to achieve high levels of thermal isolation, whilst avoiding many of the difficulties inherent in the conventional diffusive approach.

Initially we considered patterning step discontinuities into microbridges so as to produce a form of acoustic Fabry-Perot filter. However, the dominant phonon wavelengths are of order 1 μm, and so any discontinuities must be abrupt on scale sizes of say 0.2 μm. It is not possible to manufacture step changes in width with edges that are abrupt on scale sizes of significantly less than 1 μm using optical lithography. Other techniques such as Focused Ion Beam (FIB) milling are available, but we do not see how it would be possible to produce a large number of TES pixels, say in an imaging array, having near-identical characteristics. We thus conclude that Mach-Zehnder-like interferometers are preferable.

In previous work [9] we demonstrated a range of TESs where thermal isolation was limited not by diffusive scattering, but by using extremely narrow legs, thus limiting the number of elastic modes present. We also ensured ballistic transport by making the legs very short (< 4 μm). Crucially, it was possible to calculate the heat flow, the temperature dependence of the thermal conductance, and the NEP of a number of devices simply from knowing the bulk elastic constants of the supporting dielectric [9]. This work showed that we were able to manufacture and operate TESs having tiny leg geometries, that the thermal exchange with the bath was dominated by ballistic phonon transport at typical operating temperatures of ~ 100 mK, and that conductance variation due to localization was essentially eliminated. We also showed that for small cross-sections (~ 700 nm x 200 nm) the effective number of phonon modes available along each leg is close to the theoretical lower limit of 4. The devices had thermal characteristics that were entirely accounted for by ballistic elastic-mode considerations.

As the cross-section of a microbridge is reduced, the number of propagating modes eventually limits at 4 (longitudinal, in-plane flexure, out-of-plane flexure, and torsion) and our devices were close to this limit, corresponding to a limiting NEP of 1 aW Hz$^{-1/2}$. Phonon NEPs of this order are more than sufficient for use in low-noise CMB ground-based and space-borne polarimeters. The next generation of cooled-aperture far-infrared space telescopes, however, require even lower NEPs. Ideally one would like to reduce the conductance by at least an order of magnitude, whilst maintaining the many benefits of the ballistic, few-mode approach.

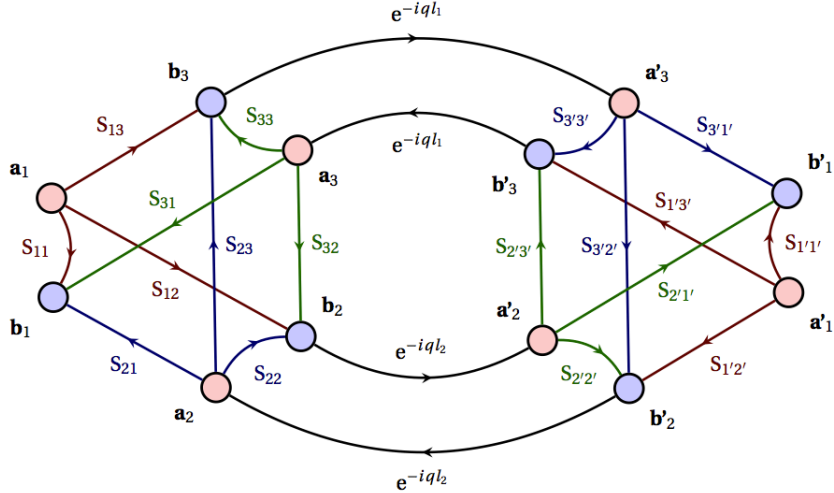

Fig 1 Directed-flow graph of an interferometer.

In order to provide a predictive tool for modelling the behavior of patterned structures, we have developed a travelling elastic-wave model of thermal transport, and applied it to simulating the behavior of micromachined elastic interferometers. In order to know the limits of what can be done, it is necessary to know the elastic and inelastic attenuation lengths. To our knowledge, there is currently no literature on the precise value of the low-temperature acoustic attenuation length in thin-film amorphous Si or $Si_xN_y$, which we conclude from our previous results must be greater than 4 μm. There is also concern that the value will depend on the stoichiometry and the presence of impurities, which are particularly difficult to control in the case of $Si_xN_y$. We know that legs that are tens of microns long show a 1/L dependence, which indicates that there must be a scattering mechanism operating on this scale size.

A central part of our recent work has been to generate a model that can take into account ballistic, diffusive and localized transport simultaneously, such that it is possible to predict the behavior of structures operating in the regime where the transport crosses over from being ballistic to diffusive. In this paper we present, for the first time, the application of the technique to elastic interferometers. We discuss some results from our elastic-wave modelling, and show how this model gives physically meaningful predictions in the ballistic/diffusive regime. It limits to ballistic interference when the interferometer is much shorter than the attenuation length, and the 1/L dependence of two parallel diffusive paths when the interferometer is much longer than the attenuation length. We then go on to show how we successfully manufactured phononic filters in the support legs of TESs.

## II. THEORY

### A. Directed-Flow Graph

The directed-flow graph of a lossless, reciprocal, Mach-Zehnder-like interferometer is shown in Fig. 1. The two power-dividing junctions (one on the left and one on the right) are assumed (although this can be relaxed) to be three-way symmetric. We also assume (although this can also be easily relaxed) that the power dividers have no spatial extent. This simplifying assumption means that differential phase changes occur only along the lengths of the two connecting arms.

The physical lengths of the waveguides in each of the two arms, $\eta$, are given by, $l_\eta$, and the transmission coefficients of these channels are simply given by the complex exponential factors shown in Fig. 1, where $q$ is the wavenumber of the propagating elastic mode. Every port, or equivalently every plane, has two nodes. The ingoing complex wave-amplitudes at a plane are denoted by a vector $\mathbf{a}$, and the outgoing complex wave-amplitudes are denoted by a vector $\mathbf{b}$. These are vectors to allow for multiple elastic modes to be accommodated. In the case of single-mode operation, or where there is no intermodal scattering, $\mathbf{a}$ and $\mathbf{b}$ become scalars. In the case of the couplers, unprimed values correspond to the left junction and primed values to the right junction. Both power dividers have three ports, denoted by the subscripted numbers $1$, $2$ and $3$. The transmission factors between every node within a divider are labeled by the scattering matrix parameters, $S_{ij}$, which denote the transmission coefficients between node $i$ and node $j$. In the case of multiple elastic modes these are block matrices. In the right divider, the subscripts are primed. Because the dividers are both the same in our case, the primed scattering parameters are identical to their unprimed counterparts. Values of S for which both subscripted indices are equal, $S_{ii}$, denote the reflection coefficient at port $i$. It can be shown that for a lossless, reciprocal, three-way symmetric junction $S_{ii}$ = -1/3 and $S_{ij}$ = 2/3, for all i,j.

### B. Calculating the Interferometer Response

For any directed-flow graph we can create a vector of all complex wave amplitudes $\mathbf{v}$, where $\mathbf{v} = [\mathbf{a_1}, ..., \mathbf{a_N}, \mathbf{b_1}, ..., \mathbf{b_N}]^T$, and where N is the total number of ports. We wish to solve for $V = \langle \mathbf{v}\mathbf{v}^\dagger \rangle$ given some power input anywhere in the flow network. $\mathbf{V}$ is a matrix of the cross-correlations between the wave-amplitudes at the nodes, and the leading diagonal of $\mathbf{V}$ gives the power flow at each node. To solve for $\mathbf{V}$, we find the transformation matrix, $\mathbf{P}$, such that $\mathbf{v} \rightarrow \mathbf{Pv}$. $\mathbf{P}$ projects the wave-amplitude at each node onto a space of its nearest dependencies. Once we have found $\mathbf{P}$, we use the formalism derived by Withington [10] to calculate $\mathbf{V}$ for some frequency-dependent blackbody thermal flux at nodes $\mathbf{a}_i$ and $\mathbf{a'}_i$. The net thermal flux across any plane, $i$, is given by $\Sigma|\langle\mathbf{a}_i\mathbf{a}_i^\dagger\rangle - \langle\mathbf{b}_i\mathbf{b}_i^\dagger\rangle|$, where the sum corresponds to a sum over elastic modes.

## C. Absorption and Re-emission

Sections A and B describe a complete model for a lossless interferometer. However, in the diffusive regime, a wave propagating along a length of microbridge in Fig. 1 will be absorbed and re-emitted, creating a temperature gradient between the power dividers. To account for absorption and phase-incoherent re-emission, we re-construct each arm in the directed-flow graph to include a large number of cells (typically 50). Each cell is essentially isothermal, and so a number of cells is needed to smoothly describe the temperature gradient. We illustrate a waveguide of this kind in Fig. 2, in which $\boldsymbol{\alpha}$ and $\boldsymbol{\beta}$ denote wave-amplitudes propagating in opposing directions at a plane. The transmission coefficient between pairs of neighbouring planes is $g\exp(-iql_n/N)$, where $N$ is the number of cells and $g$ is a unitless transmission factor < 1. The inclusion of loss simply results in 'lossy ballistic' behaviour, it does not lead to diffusive transport. To accommodate diffusion it is necessary to introduce pair-correlated sources along the whole length of the structure to model the re-radiation of phase incoherent waves by the losses in each cell. In the steady state, the total net fluxes at all (N+1) planes are equal. We then solve for the temperature each cell must have in order to reradiate a thermal flux that maintains this steady state condition. For a complete description of how this is done refer to Withington [10].

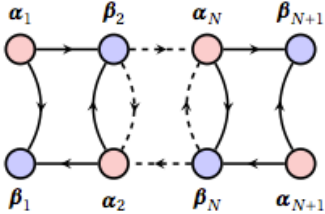

Fig. 2 A waveguide with N+1 planes.

When our thermal modelling technique is applied to an interferometer, it can be shown that the number of unknowns is equal to the number of equations, and therefore that the formulation of the problem is well conditioned. Also, in practice, we find that the technique converges rapidly and reliably even when a large number of cells is used.

## III. SIMULATION RESULTS

### D. Fabry-Perot Response and Resonant Interference

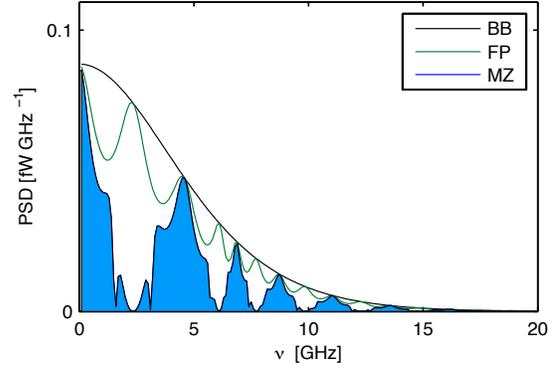

Fig 3 Net power spectral density, PSD, against phonon frequency, $\nu$, for two blackbodies exchanging power via no interferometer (black line, labelled BB), via an interferometer with equal $l_n$, where $l_1 = l_2 = 2$ μm (green line, labelled FP) and an interferometer with non-equal $l_n$, where $l_1 = 2$ μm and $l_2 = 4$ μm (blue area plot, labelled MZ). Hot and cold terminations are at 100 mK and 60 mK, respectively.

To illustrate the response of an elastic-wave interferometer, consider the transmission of only the lowest-order, longitudinal acoustic mode of a single 200 nm thick, 700 nm wide amorphous $Si_xN_y$ microbridge. Fig. 3 shows the net power spectral density, PSD, as a function of phonon frequency, $\nu$, for fully ballistic transport across a single waveguide (black line, labelled BB), an interferometer with a Fabry-Perot response, having both waveguides of equal length (green line, labelled FP) and an interferometer with a Mach-Zehnder-like response, having both waveguides of different length (blue area plot, labelled MZ). We can see that for the case in which $l_1 = l_2 = 2$ μm, there is periodic interference as a function of phonon frequency, relative to the case when there is no interferometer, even though there is no path-length difference between the two arms. This behaviour can be explained by the resonant internal reflections from the ports within the interferometer. In other words, energy is trapped in, and circulates around, the two arms. These higher-order internally reflected waves interfere in the same manner as a Fabry-Perot filter and reduce the net flux even though the arms have equal lengths. The case for which $l_1 = 2$ μm and $l_2 = 4$ μm shows a greater reduction in net thermal flux, and we see additional resonant features. These occur due to the addition of interferometric interference, and the greater number of differential path lengths within the directed-flow network when $l_1 \neq l_2$. For this example, when $l_1 = l_2 = 2$ μm, the total net flux is reduced by 20.7 per cent, and when $l_1 = 2$ μm and $l_2 = 4$ μm the total net flux is reduced by 58.9 per cent.

### E. Diffusive Transport

Fig. 4 shows the net PSD transmitted across a single waveguide (black dashed line, labelled BB) and an interferometer for which $l_1 = 2$ μm and $l_2 = 4$ μm (coloured lines), with hot and cold termination temperatures of 100 mK and 60 mK, respectively. The $g$ of every cell is calculated by setting $g^N$ for the shorter waveguide. To better illustrate the behavior of the interferometer, we again include only the lowest-order longitudinal mode of a single 200 nm thick, 700 nm wide amorphous $Si_xN_y$ microbridge. The consequence of decreasing the value of $g^N$ from a highly ballistic case, $g^N = 0.9999$, to a highly diffusive case, $g^N = 0.001$, is shown. In the ballistic limit, we see strongly resonant features, as higher order internal reflections remain phase coherent. But as $g^N$ decreases, these resonant peaks flatten, as one would expect. This occurs not simply due to the reduction in the resonator quality factor as a consequence of the losses, but also because of the introduction of phase-incoherent re-radiation by the inelastic losses. In other words, the individual arms move over into the diffusive transport regime and the interference fringes are lost.

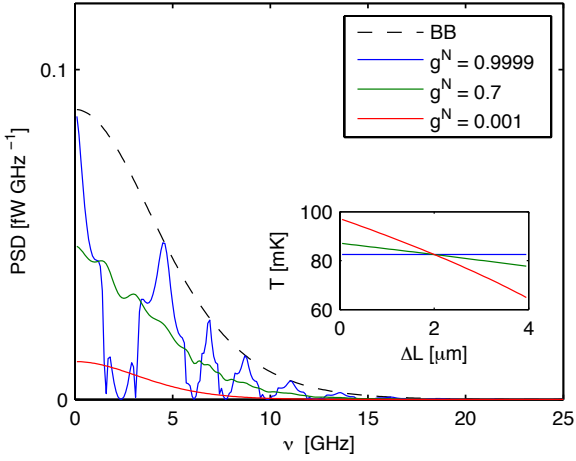

Fig 4 PSD against ν for various values of $g^N$. (Inset) Temperature, T, at different distances, L, from the hot termination

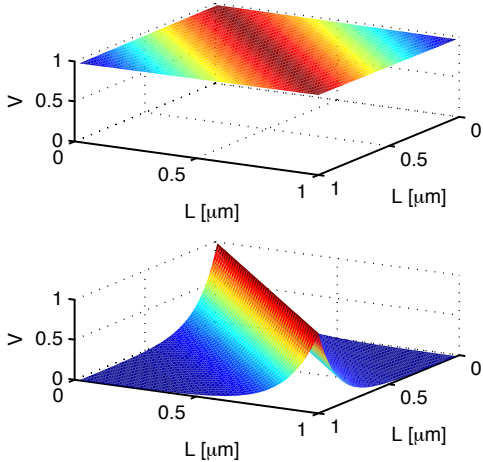

Fig 5 Normalised correlation coefficient, V, between waves travelling in the same direction at distances L = L1 and L = L2 from the hot termination. (Top) $g^N = 0.9999$ (Bottom) $g^N = 0.001$.

The inset of Fig. 4 shows the temperature of an arm as a function of position, ΔL, from the hot termination. In the ballistic limit the counter-propagating thermal fluxes are only lightly coupled to one another, and the losses thermalise everywhere at the equilibrium temperature determined by the radiation of the two thermal reservoirs. As $g^N$ decreases, the temperature becomes position-dependent, we exit the equilibrium state, and end temperatures tend towards adopting the localised hot and cold temperatures at the respective ends of the interferometer.

The transition from ballistic to highly diffusive behaviour is also illustrated in Fig. 5, which shows the normalised correlation coefficient between waves propagating in the same direction but at different planes. In the ballistic limit, the wave at every point along the waveguide is nearly perfectly correlated with the wave at every other (not perfectly, because there are two incoherent sources: the hot and cold baths). However, in the diffusive limit, the propagating waves become less correlated with points that are further away: as is expected for diffusive transport, where the waves decohere on the scale-size of the attenuation length.

### F. Thermal Filtering by Interferometers

The net thermal flux integrated across all frequencies. The net flux is transmitted from a TES of temperature $T_h = 100$ mK to a cold reservoir having temperature $T_c$. We take $T_h$ to be the critical temperature of the TES bilayer and the flux, $P$, to be the total net flux for a TES having four identical $Si_xN_y$ legs, each 200 nm thick and 700 nm wide. The plot shows the case for which there is no interferometer (blue line, labelled BB), one interferometer (green line, labelled 1 MZ) and for two interferometers (red line, labelled 2 MZ) in series per leg. Here, we model thermal transmission for the lowest six acoustic modes combined, and for interferometers that operate in the highly ballistic transport regime ($g^N = 0.9999$). For each mode we calculated its dispersion relationship using the elastic modelling technique described in our previous paper [9]. We know from this elastic-wave model that the majority of thermal power transmitted along a 700 nm x 200 nm $Si_xN_y$ microbridge is carried by the lowest 6 modes. In Fig. 6, the case having one interferometer used $l_1 = 2$ μm and $l_2 = 4$ μm. The double interferometer adds an $l_1 = 2$ μm and $l_2 = 5$ μm interferometer in-line with the other. Fitting the empirical relation $P = K(T_h^n - T^n)$ to the fluxes in Fig. 6, we determined the thermal constants $K$ and $n$, and calculated the commonly used expression for the thermal conductance, $G = nKT_h^{n-1}$ [11]. We found that the first interferometer reduces $G$ by 59 per cent and the addition of a second interferometer reduces $G$ by 74 per cent. Interestingly, the integrated flux becomes difficult to reduce below these levels because of perfect transmission at long wavelengths, where there is no interference.

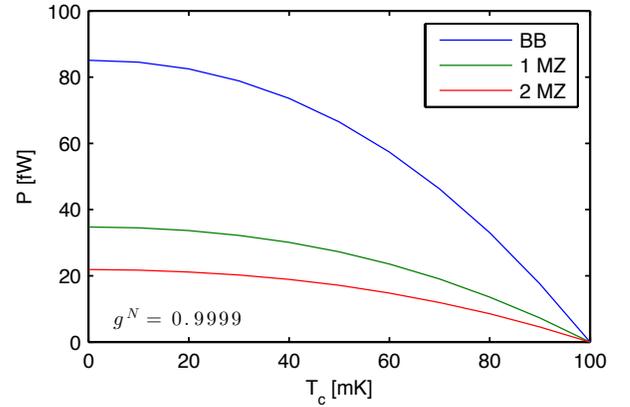

Fig 6 Integrated net thermal flux, $P$, transmitted from a TES with a critical temperature of 100 mK to a cold reservoir of temperature, $T_c$. Results are shown for no interferometer, 1 interferometer and 2 interferometers in series.

## IV. FABRICATION

A crucial question is whether it is possible to manufacture interferometric devices of this kind. Using optical lithography, we have manufactured TESs with SiN$_x$ support legs having cross-sections of 700nm x 200nm, and various configurations of patterned interferometers. Figure 7 shows an image of two of our devices. The left inset shows a leg of a device isolated by single interferometers. The right inset shows a leg of a device isolated by double interferometers. The left image has a Nb wire for biasing the TES. The bias wire has essentially no stiffness compared with the dielectric, and therefore does not change the dispersion relationships of the elastic modes. Also, because it is superconducting, it does not affect the thermal conductance. It is pleasing that we have been able to fabricate these structures, with TESs, with relative ease using optical lithography. We have also fabricated devices with varying separations between double interferometers, so as to investigate the effect of diffusively phase-decoupling two self-coherent interferometers on the same leg. Additionally, we have fabricated TESs isolated by legs that vary in length from 4 µm to 60 µm, with the aim of experimentally determining the elastic attenuation length of thin SiN$_x$. The results of experiments on these devices will be reported in due course.

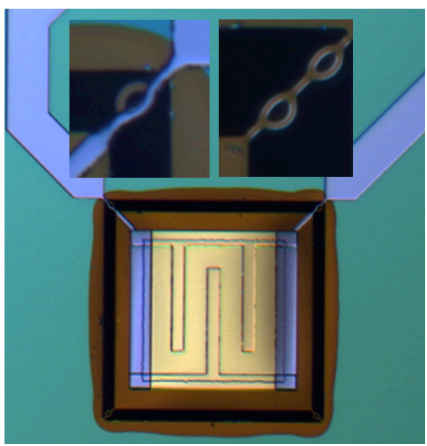

Fig 7 (Main) Image of one of our TESs with single interferometers in each leg. (Inset left) Image of a support leg with a single interferometer. (Inset right) Image of a support leg with a double interferometer.

## V. CONCLUSIONS

We have proposed the use of elastic interferometers for providing thermal isolation in low-noise Transition Edge Sensors. We believe that these structures could be potentially used for many other kinds of device, such as solid-state refrigerated platforms and thermally isolated Kinetic Inductance Detectors. We have presented a technique for calculating the behaviour of a variety of different kinds of interferometer operating in the ballistic to diffusive transport regime. Our early designs show that the thermal flux can be reduced to about 25% of the ballistic four-mode limit, which corresponds to NEPs of order $5 \times 10^{-19}$ W Hz$^{-1/2}$. We have successfully fabricated TESs having a variety interferometer designs, and have in fact shown that they are relatively straightforward to manufacture using optical lithography techniques. This work contributes to understanding how phononic structures might be used to limit heat transport in low-dimensional devices.